# Finding overlapping communities in networks by label propagation

Steve Gregory

Department of Computer Science, University of Bristol, Bristol BS8 1UB, England

**Abstract.** We propose an algorithm for finding overlapping community structure in very large networks. The algorithm is based on the label propagation technique of Raghavan, Albert, and Kumara, but is able to detect communities that overlap. Like the original algorithm, vertices have labels that propagate between neighbouring vertices so that members of a community reach a consensus on their community membership. Our main contribution is to extend the label and propagation step to include information about more than one community: each vertex can now belong to up to $v$ communities, where $v$ is the parameter of the algorithm. Our algorithm can also handle weighted and bipartite networks. Tests on an independently-designed set of benchmarks, and on real networks, show the algorithm to be highly effective in recovering overlapping communities. It is also very fast and can process very large and dense networks in a short time.

## 1. Introduction

Networks are a natural representation of various kinds of complex system, in society, biology, and other fields. Although the study of networks is not new, the amount of network data has proliferated in recent years, thanks to developments in computing and communications technology. As the number and size of network datasets has increased, so too has the interest in computational techniques that help us to understand the properties of networks.

A key property of many networks is their community structure: the tendency for vertices to be gathered into distinct groups, or *communities*, such that edges between vertices in the same community are dense but intercommunity edges are sparse. Identifying communities can allow us to understand attributes of vertices from the network topology alone. For example, all vertices in a community may be somehow related, or a vertex that appears in more than one community may play a special role. The automatic discovery of network communities can also help reveal the coarse-grained structure of networks which are too large for humans to make sense of at the level of individual vertices.

A vast number of community detection algorithms have been developed, especially in the last few years. Most of them handle unipartite networks with undirected, unweighted edges, and they vary in performance and speed. The algorithms use a wide variety of techniques, including removal of high-betweenness edges [1], optimization of modularity [2], detection of dense subgraphs [3], statistical inference [4], and many more. Even a cursory description of these algorithms is well beyond the scope of this paper. The interested reader is referred to Fortunato's excellent, comprehensive survey [5] of community detection.

The design of community detection algorithms seems to require a precise definition of *community*, but unfortunately no such definition exists [5, 6]. Each algorithm makes different assumptions consistent with the intuitive concept of a community. The majority of algorithms assume that a network contains a flat set of disjoint communities. This makes sense for many networks: for example, most employees work for a single employer, most papers are published in a single conference, etc. Some algorithms [3, 7–10, 13] allow communities to overlap, each vertex possibly appearing in more than one community. This is sometimes more realistic: for example, many researchers belong to more than one research community. Other algorithms [11–13] can find a hierarchy of communities: for example, a number of research communities each subdivided into research groups.

As network datasets become larger, the speed of community detection algorithms becomes more important. In future, algorithms will need to handle networks with millions of vertices in a reasonable time, and so any practical algorithm must have a very low time complexity.

One of the fastest algorithms proposed to date is the label propagation algorithm of Raghavan *et al* [14], which we shall call the RAK algorithm. As well as its near-linear time complexity (for sparse networks), it is very simple and has no parameters. However, like most community detection algorithms, it can detect only disjoint communities. In this paper, we propose an algorithm that generalizes the RAK algorithm to find *overlapping* communities. It takes a parameter, $v$, which controls the potential degree of overlap between communities. The RAK algorithm is essentially a special case of the proposed algorithm with $v=1$.

The next section describes the RAK algorithm. Section 3 explains and justifies the design of our algorithm, named COPRA (**C**ommunity **O**verlap **PR**opagation **A**lgorithm). In section 4 we present the results of experiments that show how our algorithm behaves, and measure its performance and speed, comparing these with other overlapping community detection algorithms. Section 5 shows how the algorithm can be simply extended to handle bipartite networks. Conclusions appear in section 6.

## 2. Detecting communities by label propagation

The RAK algorithm can be described very simply. Each vertex is associated with a label, which is an identifier such as an integer.

1. To initialize, every vertex is given a unique label.
2. Then, repeatedly, each vertex $x$ updates its label by replacing it by the label used by the greatest number of neighbours. If more than one label is used by the same maximum number of neighbours, one of them is chosen randomly. After several iterations, the same label tends to become associated with all members of a community.
3. All vertices with the same label are added to one community.

The propagation phase does not always converge to a state in which all vertices have the same label in successive iterations. To ensure that the propagation phase terminates, Raghavan *et al* propose the use of "asynchronous" updating, whereby vertex labels are updated according to the previous label of some neighbours and the *updated* label of others. Vertices are placed in some random order. $x$'s new label in the $t$th iteration is based on the labels of the neighbours that precede $x$ in the $t$th iteration and the labels of its neighbours that follow $x$ in the ($t$-1)th iteration.

The algorithm terminates when every vertex has a label that is one of those that are used by a maximum number of neighbours.

The algorithm produces groups that contain all vertices sharing the same label. These groups are not necessarily connected, in the sense that there is a path between every pair of vertices in the group passing only through vertices in the same group. Since communities are generally required to be connected, Raghavan *et al* propose a final phase that splits the groups into one or more connected communities.

The time complexity of the algorithm is almost linear in the network size. Initialization takes time $O(n)$, each iteration takes time $O(m)$, and the time for processing disconnected communities is $O(m+n)$. The number of iterations required is harder to predict, but Raghavan *et al* claim that five iterations is sufficient to classify 95% of vertices correctly.

Leung *et al* [15] have analysed the RAK algorithm in more detail. They compare asynchronous with synchronous updating, whereby the new label of each vertex in the $i$th iteration is always based on the labels of its neighbours in the ($i$-1)th iteration. They found that synchronous updating requires more iterations than asynchronous updating, but is "much more stable". They also propose restraining the propagation of labels to limit the size of communities, and a similar technique to allow detection of hierarchical communities. Both Refs. [14] and [15] hint at the possibility of detecting overlapping communities, but neither extends the algorithm to find them; we do this in the next section.

## 3. Overlapping communities

*3.1. Extending to overlapping communities*

In the RAK algorithm, a vertex label identifies a single community to which the vertex belongs. If communities overlap, each vertex may belong to more than one community. Therefore, to find overlapping communities, we clearly need to allow a vertex label to contain more than one community identifier.

As a first attempt, we might let a vertex label be a *set* of community identifiers. Initially, each vertex would be labelled by a unique identifier, and each propagation step would copy *all* of $x$'s neighbours' community identifiers into $x$'s set. This clearly will not work, because every vertex would end up labelled with the set of *all* community identifiers.

Alternatively, we could label each vertex $x$ with a set of pairs $(c,b)$, where $c$ is a community identifier and $b$ is a *belonging coefficient*, indicating the strength of $x$'s membership of community $c$, such that all belonging coefficients for $x$ sum to 1. Each propagation step would set $x$'s label to the union of its neighbours' labels, sum the belonging coefficients of the communities over all neighbours, and normalize. More precisely, assuming a function $b_t(c,x)$ that maps a vertex $x$ and community identifier $c$ to its belonging coefficient in iteration $t$,

$$b_t(c,x) = \frac{\sum_{y \in N(x)} b_{t-1}(c,y)}{|N(x)|}, \tag{1}$$

where $N(x)$ denotes the set of neighbours of $x$. We use synchronous updating, partly because it seems to give better results than asynchronous updating [15]: in our algorithm, the label of a vertex in iteration $t$ is always based on its neighbours' labels in iteration $t-1$.

Figure 1 shows the result of the first iteration using this method. After this iteration, for example, community b contains a, c, and d (with weights ¼, ½, and ⅓, respectively) and community e contains a, f, and g. However, this method is still unsuitable as a community detection algorithm because it produces as many communities as there are vertices and it converges to a solution in which all vertices have the same label: here {(a,0.248), (b,0.188), (c,0.188), (d,0.188), (e,0.188), (f,0.188)}.

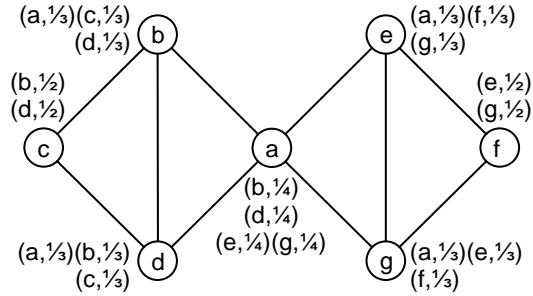

**Figure 1.** Propagation of labels: first iteration.

What is required is a way to retain more than one community identifier in each label without keeping all of them. Our method uses the belonging coefficients for this purpose: during each propagation step, we first construct the vertex label as above and then delete the pairs whose belonging coefficient is less than some threshold. We express this threshold as a reciprocal, $1/v$, where $v$ is the parameter of the algorithm. Because the belonging coefficients in each label sum to 1, $v$ represents the maximum number of communities to which any vertex can belong.

It is possible that all pairs in a vertex label have a belonging coefficient less than the threshold. If so, we retain only the pair that has the greatest belonging coefficient, and delete all others. If more than one pair has the same maximum belonging coefficient, below the threshold, we retain a randomly selected one of them. This random selection makes the algorithm nondeterministic.

After deleting pairs from the vertex label, we renormalize it by multiplying the belonging coefficient of each remaining pair by a constant so that they sum to 1.

Using this method, with *v*=2, on the above network gives the results shown in figure 2. In the first iteration, vertex c is labelled with community identifiers b and d, each with belonging coefficient ½. Because this is no less than the threshold (½), both are retained. Similarly, f is labelled with e and g. The other five vertices all have at least three neighbours, and so their belonging coefficients are all below the threshold. For example, b is labelled at first with {(a,⅓), (c,⅓), (d,⅓)}: we randomly choose c, delete a and d, and renormalize to {(c,1)}. The labels for a, d, e, and g are similarly randomly chosen. Before the final iteration, a has two neighbours labelled c and two labelled e, and so it retains both community identifiers: {(c,½), (e,½)}. The final solution therefore contains two overlapping communities: {a,b,c,d} and {a,e,f,g}.

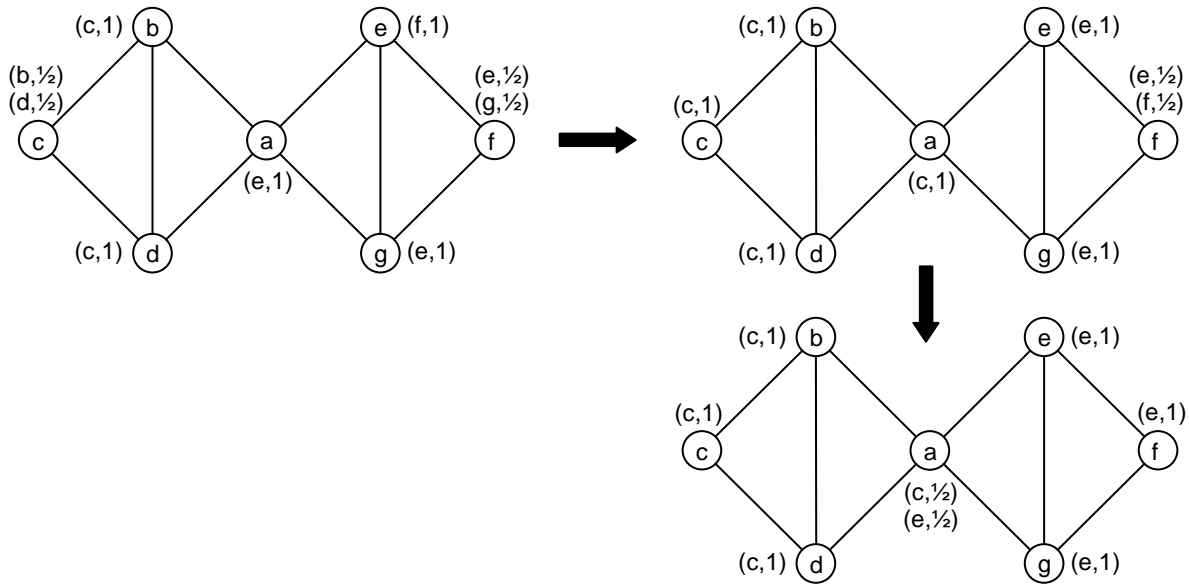

**Figure 2.** Propagation of labels with *v*=2.

Our algorithm generalizes the RAK algorithm. If *v*<2 they are essentially the same: the label of a vertex can contain only one community identifier, each propagation step retaining the identifier used by the maximum number of neighbours.

*3.2. Design alternatives*
A few variants of our propagation method were considered but rejected. We describe them briefly in this section.

One idea was to make the maximum number of communities of a vertex dependent on its degree, so that a high-degree vertex may belong to more communities than a low-degree one. This can be done by retaining community identifiers used by at least a certain *number c* of neighbours, instead of a certain *fraction* of neighbours. The threshold for vertex *x* becomes $c/d(x)$ instead of $1/v$, allowing vertex *x* to belong to up to $d(x)/c$ communities, where $d(x)$ is the degree of *x*. This did not work well, partly because low-degree vertices (which are very common) are unable to belong to more than one community. As a compromise, an alternative definition of threshold was tried: $max(1/v, c/d(x))$, but this was also less successful than our chosen one, and requires two parameters.

Second, we experimented with alternative ways to handle the situation where all community identifiers have belonging coefficients below the threshold. In COPRA we keep *one* of these, but instead we could keep *all* of those that have the maximum value *or* none of them, leaving the vertex unlabelled. Both of these were tried but were unsuccessful.

Finally, we tried a simple method to prevent community identifiers propagating too far and forming excessively large communities; this is a potential problem with our propagation method, inherited from the RAK algorithm. Our technique checks whether, after a propagation step, vertex *x* is still among the community identifiers labelling vertex *x*. If so, we set the label to *x* only, deleting all other community identifiers from *x*'s label. This seems reasonable because label *x* has appeared on *x* more than once, and it clearly reduces the size of other communities because they no longer include *x*.

This technique was successful in improving the worst results but the best results deteriorated, probably because community sizes were sometimes limited more than they should have been. Leung et al. [15] proposed a more complicated "hop attenuation" strategy to achieve a similar effect, also with inconclusive results.

*3.3. Termination*

Like the RAK algorithm, our algorithm often does not converge to a state in which vertex labels remain constant between iterations. Community identifiers may continue to propagate indefinitely (for example, around a four-cycle). Ref. [14] solves this by a termination criterion that is unsuitable for COPRA because our vertex labels are more complex and we use synchronous updating. We therefore need a new termination condition.

In our algorithm, the number of community identifiers is initially $n$ (the number of vertices) and decreases monotonically as identifiers are deleted from the vertex labels. Eventually this number reaches a minimum, even if this is 1 (a trivial solution). That is, the set of community identifiers in use at the $t$th iteration,

$$i_t = \{c \in V : \exists x \in V (b_t(c,x) > 0)\}, \qquad (2)$$

where $V$ is the set of all vertices, reaches a minimum size.

The propagation should not terminate as soon as $i_t = i_{t-1}$, unless $|i_t| = 1$, because $i_t$ may reduce again after many more propagation steps. Even if it does not, the solution might improve further as community identifiers continue to move between vertices. Instead, we could check the *number* of vertices labelled with each community identifier,

$$c_t = \{(c,i) : c \in V \wedge i = \sum_{x \in V, b_t(c,x) > 0} 1 \}, \qquad (3)$$

but this usually changes in successive iterations, so we cannot expect that $c_t = c_{t-1}$.

Our chosen method is to compute the *minimum* number of vertices labelled with each community identifier since the number of identifiers last reduced:

$$\begin{aligned} m_t &= \{(c,i) : \exists p \exists q ((c,p) \in c_{t-1} \wedge (c,q) \in c_t \wedge i = \min(p,q))\} \text{ if } i_t = i_{t-1} \\ m_t &= c_t \text{ otherwise} \end{aligned} \qquad (4)$$

We stop the propagation as soon as $m_t = m_{t-1}$. It is easy to see that this will happen after a finite number of steps, and so the algorithm is guaranteed to terminate.

Although we cannot prove that the $t$th iteration always represents the best solution, our termination criterion works well in practice and is inexpensive to compute. We use it for all results presented in section 4.

*3.4. Postprocessing*

After termination, each vertex whose label contains community identifier $c$ is simply allocated to community $c$. However, this method might form communities that are subsets of (or identical to) others. We avoid this by keeping track of the subset relationship between communities as they are constructed, and then deleting any community that is a subset of any other. This is faster than checking for set inclusion after all communities are constructed: it can be done in linear time.

As with the RAK algorithm, the communities found may not be connected, and so finally we split all disconnected communities into smaller, connected ones.

*3.5. The COPRA algorithm*

The complete COPRA algorithm is shown in figures 3 and 4. It keeps two vectors of vertex labels: *old* and *new*; *old.x* (resp. *new.x*) denotes the previous (resp. updated) label for vertex $x$. Each vertex label is a set of pairs $(c,b)$, where $c$ is a community identifier and $b$ is the belonging coefficient. $N(x)$ is the set of neighbours of vertex $x$.

*3.6. Complexity*

The time complexity of each step of the algorithm is estimated below. $n$ is the number of vertices, $m$ is the number of edges, and $v$ is the parameter (maximum number of communities per vertex).

1. Initialization takes time $O(n)$.
2. This phase constructs an updated label (maximum size $O(vm/n)$) for each of the $n$ vertices. For each vertex $x$, it iterates through the $O(m/n)$ neighbours. For each neighbour $y$, it iterates through all (at most $v$) community identifiers in $y$'s label, looking up each one in $x$'s updated label (which takes time $O(\log(vm/n))$ if a tree data structure is used). The normalization and deletion activities for each vertex $x$ iterate through the updated label of $x$ (time $O(vm/n)$ per vertex). The time for the whole phase is therefore $O(vm \log(vm/n))$.
3. This iterates through all $O(v)$ community identifiers of each of the $n$ vertices, looking up each in a set of size $O(n)$. Using a hash table data structure, the total time is $O(vn)$.
4. Similar to phase 3. The total time is $O(vn)$.
5. Iterates through all community identifiers, $c$, of all vertices, $x$ ($O(vn)$ steps). At each step, it adds $x$ to community $c$ and updates the set of communities of which $c$ is a subset (time $O(v^2)$, because the set sizes are at most $v$). The total time is $O(v^3 n)$.
6. Iterates through at most $n$ communities: time $O(n)$.
7. This is similar to the RAK algorithm but there are up to $vn$, instead of $n$, vertices in the communities at this stage, and so the total time is $O(v(m+n))$.

Phases 2-4 are repeated, so the time per iteration is $O(vm \log(vm/n))$. The initial and final steps take time $O(v(m+n)+v^3 n)$.

For a sparse network, the time complexity is therefore $O(v^3 n)$ plus $O(vn \log(v))$ per iteration. Assuming that $v$ is a small integer the execution time per iteration is essentially linear.

*3.7. Weighted networks*

To extend COPRA to weighted networks, we simply replace the line $b \leftarrow b_y$, in the *Propagate* operation, by $b \leftarrow b_y w_{xy}$, where $w_{xy}$ is the weight of the edge $\{x,y\}$.

1. For each vertex $x$:
   $old.x \leftarrow \{(x,1)\}$.
2. For each vertex $x$:
   *Propagate*($x$, *old*, *new*).
3. If $id(old) = id(new)$:
   $min \leftarrow mc(min, count(new))$.
   Else:
   $min \leftarrow count(new)$.
4. If $min \neq oldmin$:
   $old \leftarrow new$.
   $oldmin \leftarrow min$.
   Repeat from step 2.
5. For each vertex $x$:
   $ids \leftarrow id(old.x)$.
   For each $c$ in $ids$:
     If, for some $g$, $(c,g)$ is in *coms*, $(c,i)$ in *sub*:
       $coms \leftarrow coms - \{(c,g)\} \cup \{(c,g\cup\{x\})\}$.
       $sub \leftarrow sub - \{(c,i)\} \cup \{(c,i\cap ids)\}$.
     Else:
       $coms \leftarrow coms \cup \{(c,\{x\})\}$.
       $sub \leftarrow sub \cup \{(c,ids)\}$.
6. For each $(c,i)$ in *sub*:
   If $i \neq \{\}$: $coms \leftarrow coms - (c,g)$.
7. Split disconnected communities in *coms*.

**Figure 3.** The COPRA algorithm.

```
Propagate(x, source, dest):
    dest.x ← {}.
    For each y in N(x):
        For each (c,b_y) in source.y:
            b ← b_y.
            If, for some b_x, (c,b_x) is in dest.x:
                dest.x ← dest.x − {(c,b_x)} ∪ {(c,b_x+b)}.
            Else: dest.x ← dest.x ∪ {(c,b)}.
    Normalize(dest.x).
    b_max = 0.
    For each (c,b) in dest.x:
        If b < 1/v:
            dest.x ← dest.x − {(c,b)}.
            If b > b_max:
                b_max ← b.
                c_max ← c.
    If dest.x = {}: dest.x ← {(c_max,1)}.
    Else: Normalize(dest.x).

Normalize(l):
    sum ← 0.
    For each (c,b) in l: sum ← sum + b.
    For each (c,b) in l: l ← l − {(c,b)} ∪ {(c,b/sum)}.

id(l):
    ids ← {}.
    For each l.x in l: ids ← ids ∪ id(l.x).
    Return ids.

id(x):
    ids ← {}.
    For each (c,b) in x: ids ← ids ∪ {c}.
    Return ids.

count(l):
    counts ← {}.
    For each l.x in l:
        For each (c,b) in l.x:
            If, for some n, (c,n) is in counts:
                counts ← counts − {(c,n)} ∪ {(c,n+1)}.
            Else: counts ← counts ∪ {(c,1)}.
    Return counts.

mc(cs_1, cs_2):
    cs ← {}.
    For each (c,n_1) in cs_1, (c,n_2) in cs_2:
        cs ← cs ∪ {(c, min(n_1,n_2))}.
    Return cs.
```

**Figure 4.** The COPRA algorithm (contd.).

## 4. Experiments

*4.1. Methodology*

There are two ways to evaluate the performance of a community detection algorithm. One is to run the algorithm on real-world network data. A problem with this is that it is hard to judge the communities found because we usually do not know the real communities that are present in the original data; even if we do, they are not necessarily reflected in the network structure. The other method is to use randomly-generated synthetic networks based on a known community structure and compare the known communities with those found by the algorithm. A benefit of this method is that we can vary the parameters of the networks to analyse the algorithm's behaviour in detail. The drawback is that artificial networks might not share the properties of real networks.

For synthetic networks, there are various standard measures [16–18] that can be used to compare the known and computed communities, but none of these are suitable for overlapping communities. Exceptions are the Omega index [19], which is based on the Adjusted Rand index, and a new variant of the Mutual Information measure [13], extended to handle overlapping communities. We use the Normalized Mutual Information (NMI) measure of Ref. [13] in the experiments reported in this paper.

For real networks, the quality of a solution is usually assessed by the relative density of edges *within* communities and *between* communities. The most common measure of this is modularity [20, 21]: a high value indicates more intracommunity edges than would be expected by chance. The original modularity measure is defined only for disjoint communities, but Nicosia *et al* [22] have recently designed a variant that is defined also for overlapping communities. We use this *overlap modularity* ($Q_{ov}$) measure for the experiments in this paper. Its value depends on the number of communities to which each vertex belongs and the strength of its membership to each community; we assume that each vertex belongs equally to all of the communities of which it is a member.

The overlap modularity measure is defined in terms of a function *f*, which in turn is defined as:

$$f(x) = 2px - p \ , \tag{5}$$

but the value of *p* is not specified in Ref. [22]. Therefore, we use the following definition of *f*, suggested in Ref. [23]:

$$f(x) = 60x - 30 \ . \tag{6}$$

As well as measures of solution quality, we measure:

1. The number of non-singleton communities. This is more reliable than the total number of communities because some algorithms assign *every* vertex to at least one community even if it has only one member, while others leave some vertices unassigned to any community.
2. The overlap: the average number of communities to which each vertex belongs. This is the sum of the sizes of *all* communities (including singletons) divided by the number of vertices, *n*.
3. The execution time. This was measured using the programs running under Linux on an AMD Opteron 250 CPU at 2.4GHz.

This section describes experiments on both artificial and real networks. For the former, we use the network generator of Lancichinetti *et al* [25]. This produces benchmark networks which are claimed to possess properties found in real networks, such as heterogeneous distributions of degree and community size. Although not described in Ref. [25], the generator also allows communities to overlap, with the restriction that every overlapping vertex belongs to a fixed number of communities.

The parameters of the benchmark networks are: the number of vertices (*N*), the average degree ($\langle k \rangle$), the mixing parameter ($\mu$ – each vertex shares a fraction $\mu$ of its edges with vertices in other communities), the minimum community size ($c_{min}$), and the number of vertices that are in more than one community ($o_n$).

Like the experiments reported in Ref. [26], we keep the remaining parameters fixed for all of our experiments: the maximum degree $k_{max}$ is 2.5*$\langle k \rangle$, the maximum community size $c_{max}$ is 5*$c_{min}$, and each overlapping vertex belongs to two communities ($o_m$=2). The exponents of the power-law distribution of vertex degrees ($\tau_1$) and community sizes ($\tau_2$) are -2 and -1, respectively.

Results are obtained by running COPRA once on each of 100 random networks (unless stated otherwise) generated using the same parameters, and averaging the results. For real networks, we run COPRA 100 times on each network and average the results, since the algorithm can be rather nondeterministic. The real networks that we use, and their sources and sizes, are listed in columns 1-4 of table 1. Most of these are social networks, while "protein-protein" is a biological network and "word_association" is a network of word meanings.

**Table 1.** Real networks used and results of COPRA.

| Name | Ref | Vertices | Edges | $v$ | Mean $Q_{ov}$ | Std dev $Q_{ov}$ | # non-singleton communities | Overlap | # iterations | Execution time (s) |
|---|---|---|---|---|---|---|---|---|---|---|
| netscience | [27] | 379 | 914 | 5 | 0.812 | 0.026 | 33.2 | 1.193 | 24.3 | 0.04 |
| jazz | [28] | 198 | 2742 | 2 | 0.716 | 0.052 | 2.8 | 1.004 | 9.0 | 0.03 |
| email | [29] | 1133 | 5451 | 2 | 0.506 | 0.237 | 8.5 | 1.024 | 35.7 | 0.22 |
| protein-protein | [3] | 2445 | 6265 | 5 | 0.552 | 0.066 | 135.6 | 1.176 | 56.2 | 0.75 |
| blogs | [8] | 3982 | 6803 | 9 | 0.748 | 0.007 | 133.6 | 1.168 | 166.8 | 2.9 |
| PGP | [30] | 10680 | 24316 | 11 | 0.788 | 0.017 | 258.7 | 1.186 | 249.6 | 14.4 |
| word_association | [3] | 7205 | 31784 | – | ~0 | – | – | – | – | – |
| blogs2 | [8] | 30557 | 82301 | 2 | 0.604 | 0.032 | 851.8 | 1.027 | 172.5 | 31.7 |
| cond-mat-2003 | [31] | 27519 | 116181 | 1 | 0.675 | 0.057 | 1788.0 | 1.000 | 331.4 | 66.5 |

We shall henceforth refer to the overlap modularity measure as simply *modularity* or $Q_{ov}$. We sometimes use the term *solution* to refer to the *partition* (division of a network into disjoint communities) or *cover* (division into overlapping communities) computed by an algorithm. Although modularity is a function of a cover and a network, we sometimes refer to the "modularity of algorithm $A$ on network $N$", meaning the modularity of the cover produced by algorithm $A$ when run on network $N$. Similarly, by "NMI of algorithm $A$ on network $N$" we mean the NMI that quantifies the difference between the cover produced by algorithm $A$ and the communities used to construct $N$.

*4.2. Properties of the algorithm*
Here we show the results of experiments designed to evaluate the behaviour of the COPRA algorithm and test the effect of varying the $v$ parameter. This reveals the main differences between COPRA and the RAK algorithm, which corresponds to $v=1$.

First, the COPRA algorithm was run on the PGP network, and allowed to run continuously, without its termination check. As figure 5 shows, for each $v$, the modularity of the solution increases with the number of iterations, reaches a maximum, and then fluctuates slightly. The figure shows the point at which the algorithm normally terminates, using our termination condition (section 3.3). This invariably happens soon after the maximum modularity has been reached, showing that our termination criterion is very effective, at least on this network.

Another property that stands out from figure 5 is the strong influence that the value of $v$ has on modularity. The modularity increases as $v$ varies from 2 to 9, but $v=1$ gives a solution better than $v=2$ or 3. This is also shown in figure 6. The highest modularity is obtained when $v$ is 9, 10, or 11, beyond which it declines gradually.

Figure 7 shows results of running COPRA on two synthetic networks: one with disjoint communities and one with overlapping communities. The *y*-axis shows the NMI, which measures the closeness of the found communities to the real communities. For the disjoint network ($o_n=0$, $\mu=0.4$), $v=1$ finds a near-perfect solution. More surprisingly, $v$ can be increased to 9 with no adverse effect, with the best solution given by $v$ between 4 and 9. For the overlapping network ($o_n=2500$, $\mu=0.1$), the overlap (average number of communities per vertex) is 1.5. This time, $v=1$ gives a poor solution, as expected. The solution quality improves as $v$ is increased, achieving an almost perfect solution when $v$ is 7 or 8.

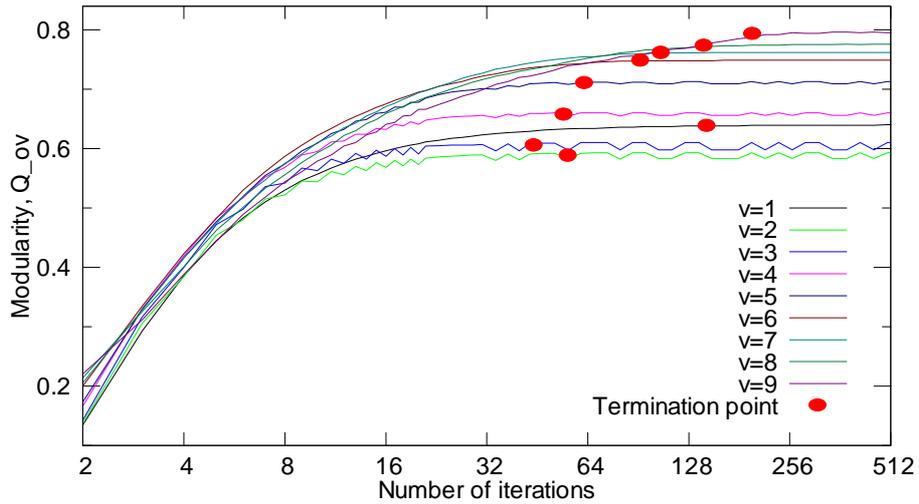

**Figure 5.** Modularity and termination points of COPRA on PGP network.

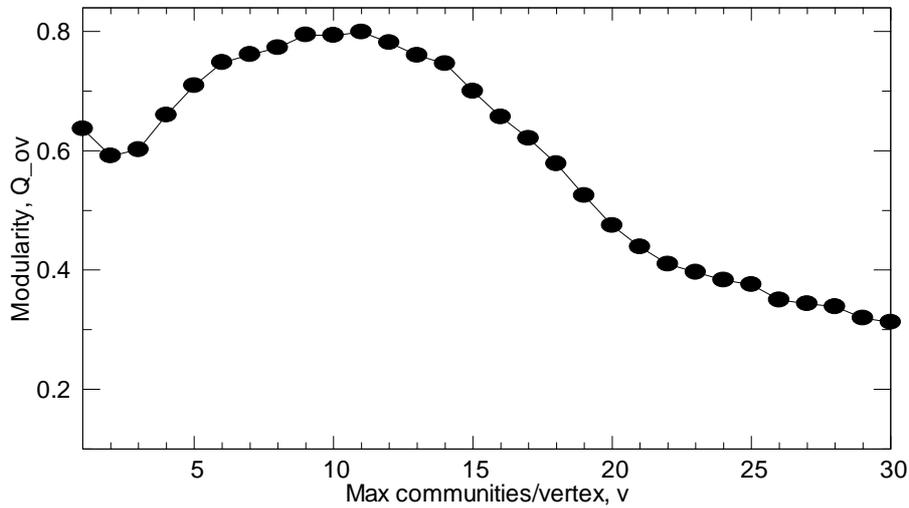

**Figure 6.** Modularity of COPRA on PGP network.

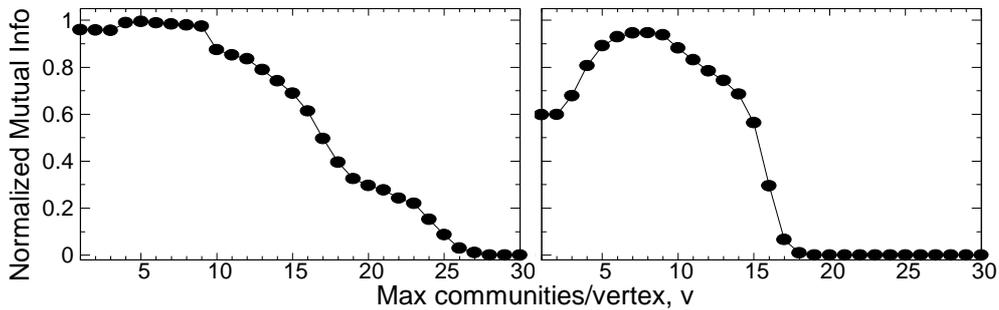

**Figure 7.** NMI of COPRA on synthetic networks. Left: Disjoint communities: $\mu=0.4$, $o_n=0$. Right: Overlapping communities: $\mu=0.1$, $o_n=2500$. The other parameters are: $N=5000$, $\langle k \rangle=20$, and $c_{min}=10$.

To understand why $v$ affects the solution quality, we plot some properties of the COPRA solution: the number of communities found and the amount of overlap between them. Figure 8 shows these for the two artificial networks. For the disjoint one, the known solution contains, on average, 203 disjoint communities (overlap=1). The correct overlap is obtained by $v=1,…,9$, while $v=4,…,13$ finds an almost perfect number of communities. This explains why $v=4,…,9$ finds the solution with the maximum NMI. For the overlapping network, as $v$ increases from 1 to 7, the number of communities found increases slightly from 275 to 285 (the correct number is 305). The overlap also increases with $v$, but in the range $v$=6,…,9, it varies only between 1.46 and 1.51, very close to the correct value of 1.5. This is why $v$=7 or 8 finds the best solution.

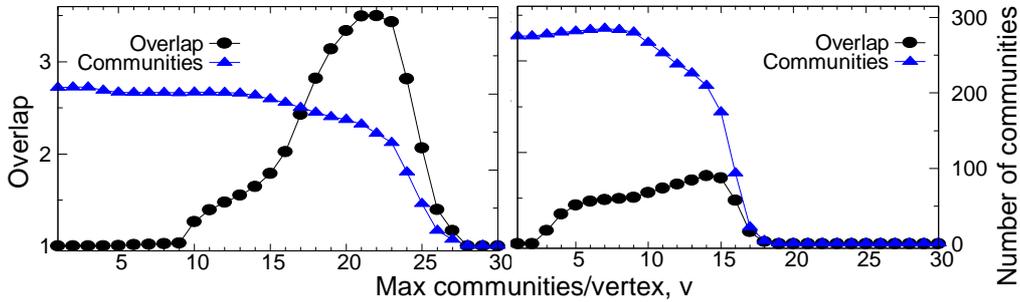

**Figure 8.** Number of communities found by COPRA and the overlap between them. Left: Synthetic network with disjoint communities: $\mu$=0.4, $o_n$=0. Right: Synthetic network with overlapping communities: $\mu$=0.1, $o_n$=2500. The other parameters are $N$=5000, $\langle k \rangle$=20, and $c_{min}$=10.

Finally, we experimentally evaluate the speed of COPRA. This depends on both the execution time per iteration and the number of iterations. In section 3.6, we predicted that the worst-case time per iteration is $O(vm \log (vm/n))$, but did not attempt to predict the number of iterations required. Figure 9 shows the timing results for our two example networks for various values of $v$. The time per iteration increases with $v$, but the increase is sublinear and much better than predicted. This is because most vertices contain far fewer than $v$ community identifiers. The number of iterations seems to be smaller for the value of $v$ that gives the correct solution. When $v$ is too high, the amount of overlap becomes excessive and so does the number of iterations.

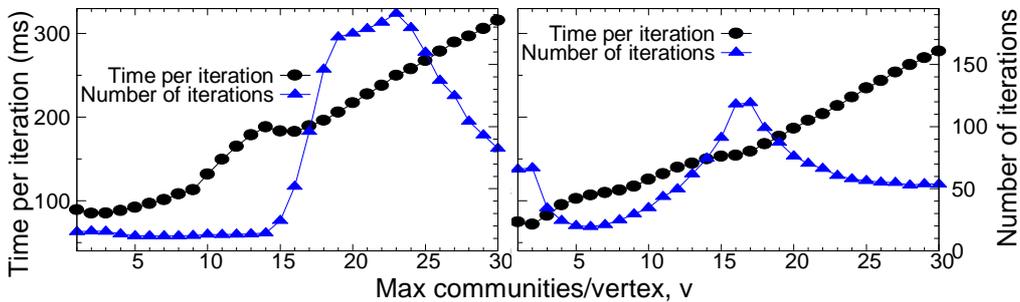

**Figure 9.** Time per iteration and number of iterations for COPRA. Left: Synthetic network with disjoint communities: $\mu$=0.4, $o_n$=0. Right: Synthetic network with overlapping communities: $\mu$=0.1, $o_n$=2500. The other parameters are $N$=5000, $\langle k \rangle$=20, and $c_{min}$=10.

Figure 10 shows the effect of varying $m$, the number of edges. There are two ways to do this: by varying the number of vertices *or* the (average and maximum) degree. When we increase the number of vertices, the time per iteration increases linearly with $m$. The number of iterations also increases, but sublinearly. When we increase only the degree, the time per iteration again increases linearly. More interestingly, the number of iterations *decreases* as the degree increases.

Results of COPRA on our real networks are shown in columns 5-11 of table 1. Column 5 shows the value of the $v$ parameter that was found to maximize the modularity of the solution. COPRA was then run 100 times with the chosen $v$. The mean and standard deviation of modularity are shown in the table, together with the average number of (non-singleton) communities found, overlap, number of iterations required, and total execution time.

Considering that the algorithm can be highly nondeterministic, the results are mostly very stable. The least stable network is "email", with a standard deviation of 0.237. Results are worst for the "word_association" network, for which COPRA finds only a single "giant" community.

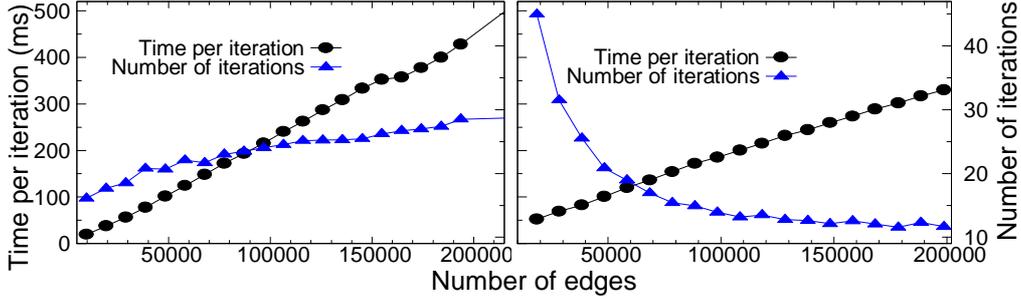

**Figure 10.** Time per iteration and number of iterations for COPRA on synthetic networks. Left: Varying number of vertices: $N$=1000-20000, $\langle k \rangle$=20. Right: Varying average degree: $\langle k \rangle$=8-80, $N$=5000. The other parameters are $\mu$=0.1, $c_{min}$=$\langle k \rangle/2$, and $o_n$=$N/2$.

*4.3. Comparison with other algorithms*

In this section, we use synthetic networks to compare the performance of COPRA with some other algorithms. There are infinitely many networks that could be used as benchmarks for comparison. To avoid any risk of bias, we use the same ones as Lancichinetti and Fortunato used in their experimental analysis of several community detection algorithms [26].

The first set of benchmarks used in Ref. [26] contains disjoint communities only. The network size is either 1000 or 5000, community sizes are in the range 10-50 or 20-100, the mixing parameter $\mu$ varies from 0.1 to 0.8, and other parameters are fixed.

Figure 11(a) shows the results (averaged over 10 runs) of the CFinder algorithm [3, 24, 32], which finds communities that are composed of adjacent $k$-cliques; $k$ is the parameter of the algorithm. Results (for $k$=4, which performs best) show that CFinder does well up to $\mu$=0.4 but only for small communities. Figure 11(b) shows analogous results from the LFM algorithm of Lancichinetti *et al* [13]. This performs better than CFinder up to $\mu$=0.5 or 0.55, but then deteriorates suddenly.

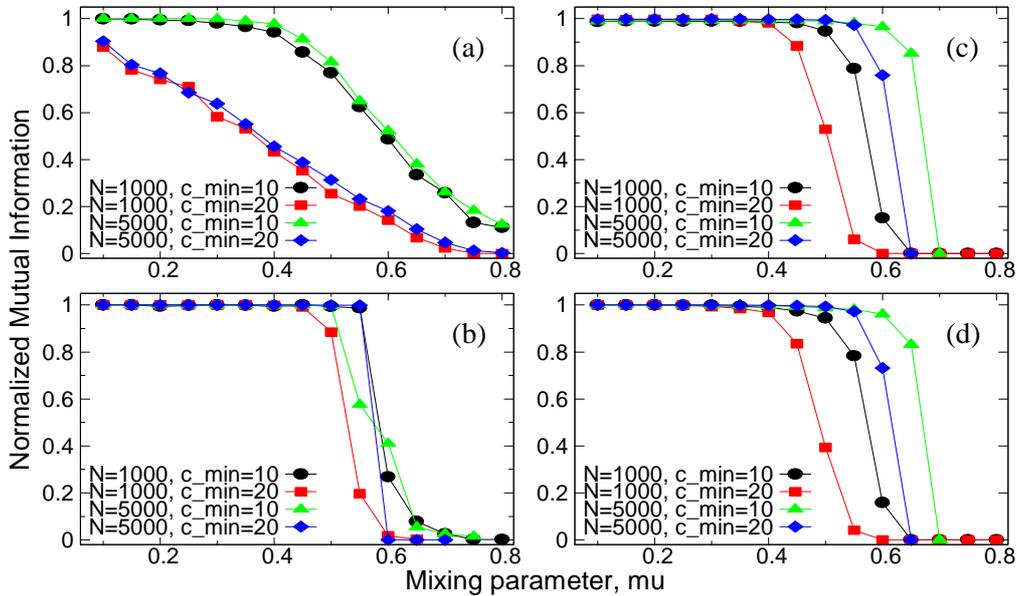

**Figure 11.** NMI of algorithms on synthetic network with disjoint communities: $\langle k \rangle$=20, $o_n$=0. (a) CFinder, $k$=4. (b) LFM. (c) COPRA, $v$=1. (d) COPRA, $v$=4.

The corresponding COPRA results are also shown in figure 11. In figure 11(c), $v$=1, and so COPRA is identical to the RAK algorithm, except for the synchronous updating and the termination criterion. The results are excellent up to $\mu$=0.45-0.5 for small networks and 0.55-0.6 for large networks, beyond which they deteriorate sharply. Although these results are better than CFinder's,

COPRA benefits from being restricted to disjoint communities, because $v$=1. Therefore, in figure 11(d), we remove this advantage by increasing $v$ to 4. COPRA still does almost equally well. All three algorithms perform well, given that they are not specialized for disjoint communities *and* have to discover the number of communities without user input.

The next class of benchmarks was designed [26] to test CFinder's ability to detect overlapping communities. This time, the fraction of overlapping vertices ($o_n/N$) varies while the mixing parameter $\mu$ is set to either 0.1 or 0.3. As before, the network size is either 1000 or 5000 and community sizes are in the range 10-50 or 20-100, and so there are eight sets of benchmarks. Results are plotted in figures 12-15. The left side of each figure shows the results from CFinder (for $k$=4) as well as LFM [13] and CONGO [9, 33] (for $h$=2 and $h$=3). The right side shows the results from COPRA for several values of $v$. For clarity, we omit $v$=2 because its results are close to $v$=1.

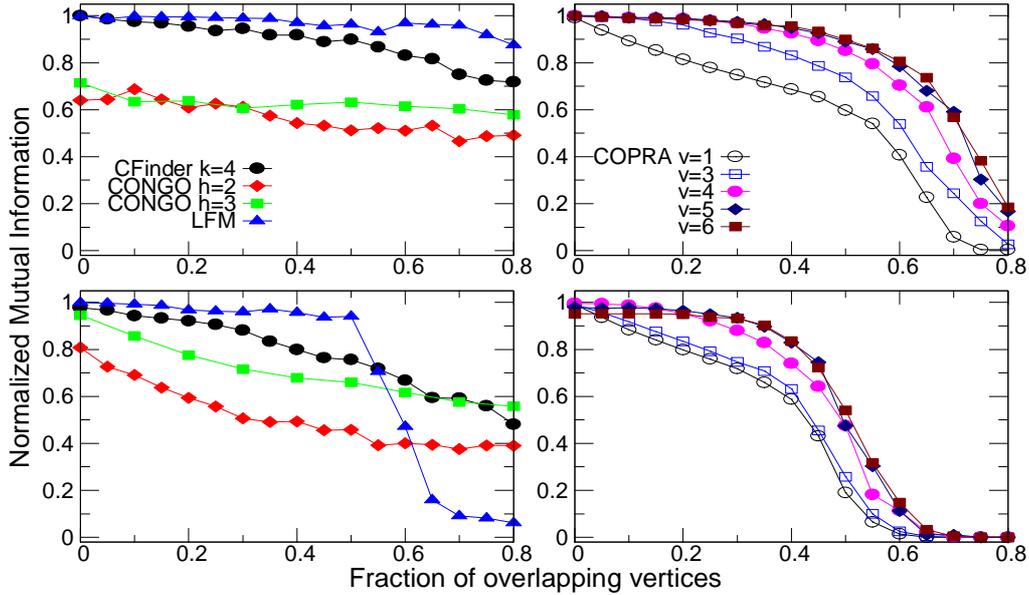

**Figure 12.** NMI of CFinder and CONGO (left) and COPRA (right) on small network with small communities: $N$=1000, $\langle k \rangle$=20, $c_{min}$=10. Top: $\mu$=0.1. Bottom: $\mu$=0.3.

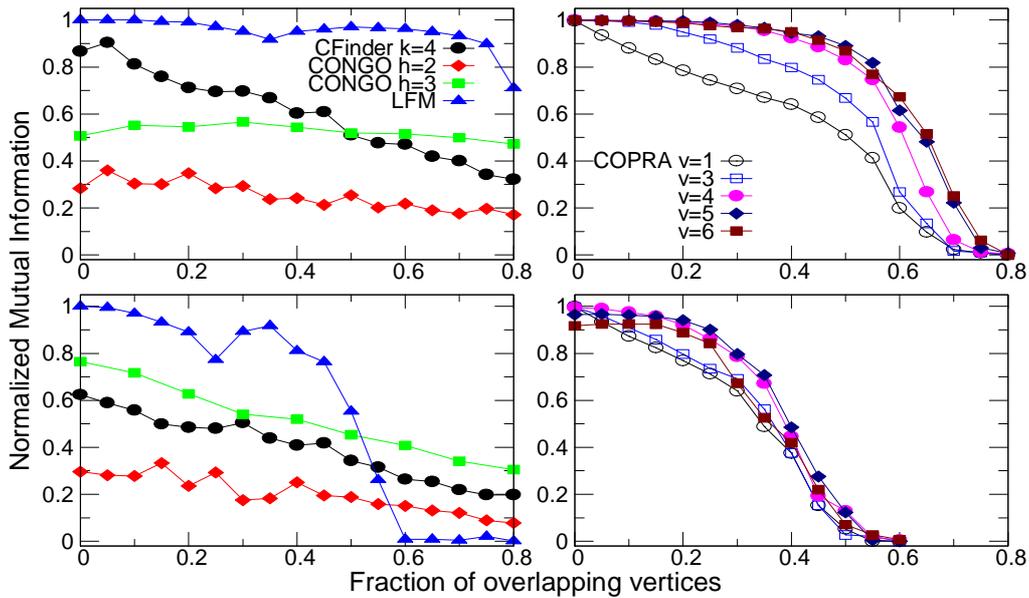

**Figure 13.** NMI of CFinder and CONGO (left) and COPRA (right) on small network with large communities: $N$=1000, $\langle k \rangle$=20, $c_{min}$=20. Top: $\mu$=0.1. Bottom: $\mu$=0.3.

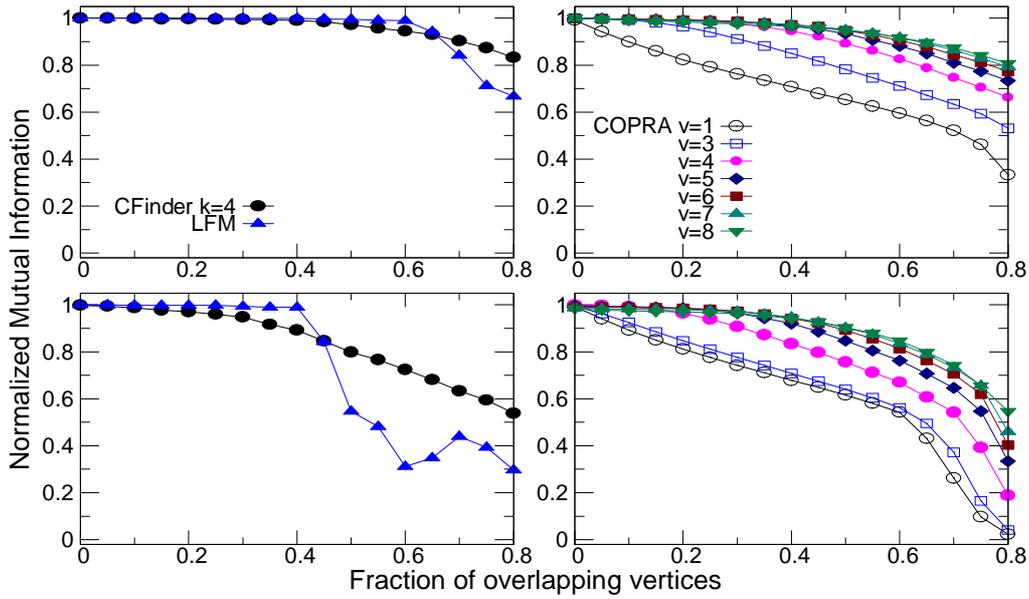

**Figure 14.** NMI of CFinder (left) and COPRA (right) on large network with small communities: $N$=5000, $\langle k \rangle$=20, $c_{min}$=10. Top: $\mu$=0.1. Bottom: $\mu$=0.3.

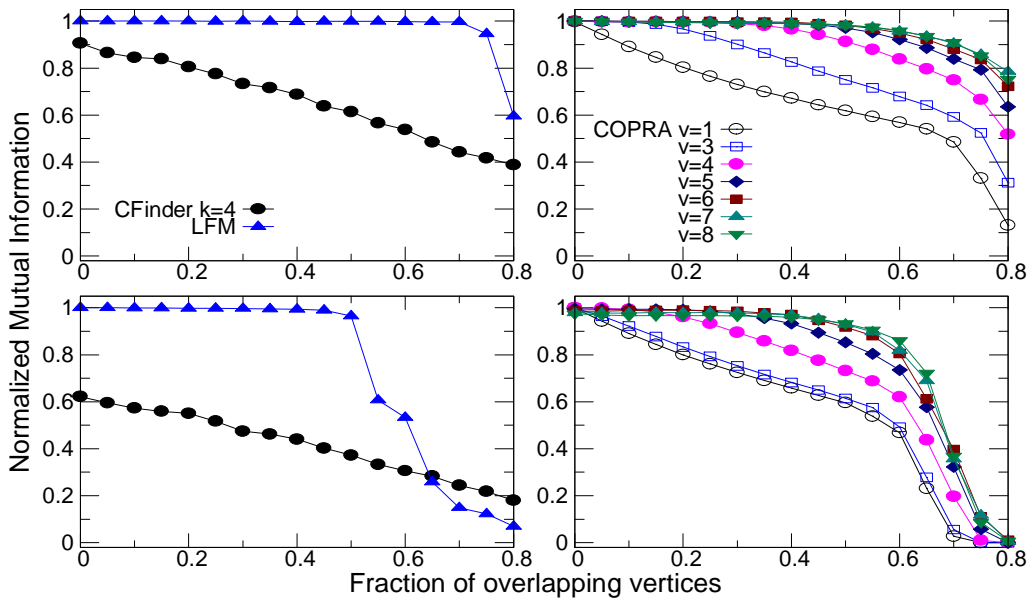

**Figure 15.** NMI of CFinder (left) and COPRA (right) on large network with large communities: $N$=5000, $\langle k \rangle$=20, $c_{min}$=20. Top: $\mu$=0.1. Bottom: $\mu$=0.3.

Like figure 11, these results show that CFinder is more effective for smaller communities while COPRA gives its best results for large networks. Generally, the COPRA results for $v>1$ are much better than for $v=1$, because the latter can only find disjoint communities. The optimal value of $v$ varies: $v=6$ works best for small networks while $v=8$ is best for large networks. Compared with CFinder and CONGO, COPRA usually performs much better when overlap is relatively low, but declines more sharply as overlap increases too far. The performance of LFM is broadly similar to that of COPRA, but slightly better than COPRA when mixing ($\mu$) is low.

Figure 16 shows the overall execution time for CFinder, CONGO, LFM, and COPRA for synthetic networks of different sizes. These results are averages of 10 runs. First, the (average and maximum) degree is fixed, while the number of vertices is varied. The largest networks that the programs could handle in 25 minutes have 100000 edges (CFinder), 200000 edges (CONGO and LFM), and 7000000 edges (COPRA). CONGO's time increases linearly; COPRA is much faster but its time increases slightly worse than linearly, because of the (slowly) increasing number of iterations. Next, the degree is varied while the number of vertices remains constant. This time, the COPRA execution time increases sublinearly with the number of edges, because of the reduced number of iterations. In contrast, the time for both CFinder and CONGO increase rapidly with size. LFM is faster but still much slower than COPRA.

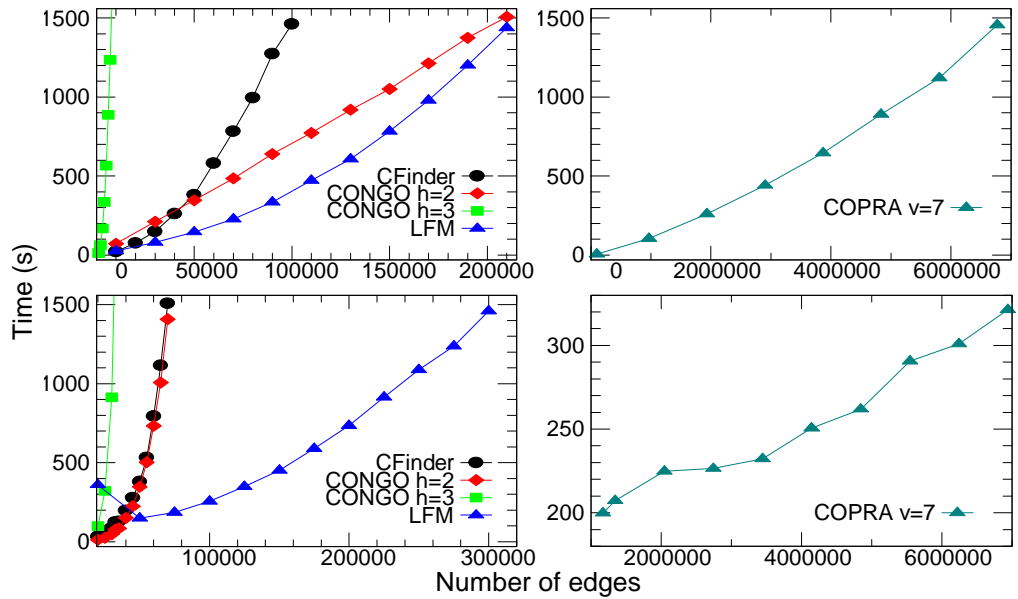

**Figure 16.** Execution time of CFinder and CONGO (left) and COPRA (right) on synthetic networks of different sizes. Top: $N$=1000-70000, $\langle k \rangle$=20. Bottom left: $\langle k \rangle$=8-28, $N$=5000. Bottom right: $\langle k \rangle$=14-80, $N$=175000. The other parameters are $\mu$=0.1, $c_{min}$=$\langle k \rangle$/2, $o_n$=$N$/2.

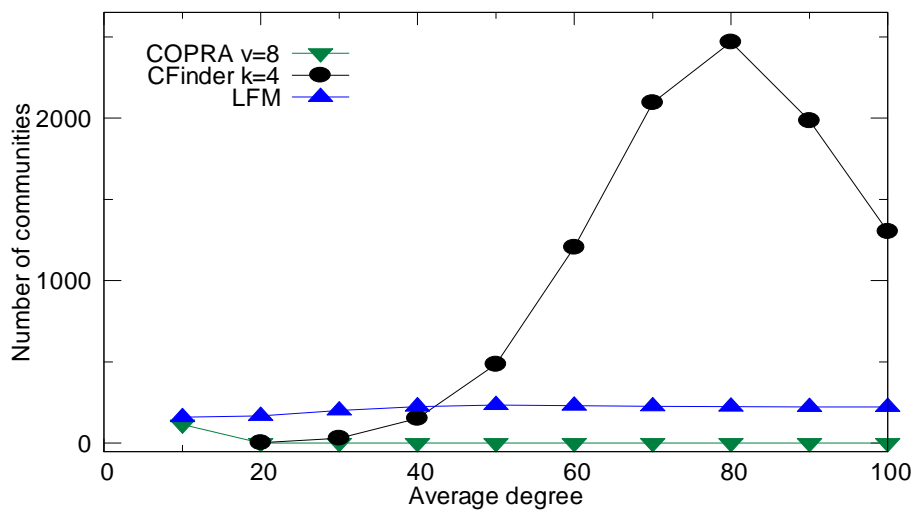

**Figure 17.** Results of COPRA, CFinder, and LFM algorithms on synthetic Erdös-Rényi random networks with 1000 vertices.

Lancichinetti and Fortunato [26] point out that a good community detection algorithm should not only find communities when they exist, but should also avoid finding spurious "pseudocommunities" in a network with no intrinsic community structure, such as a random network. On such a network, the correct solution should be a trivial partition that contains only one community or only singleton communities. In figure 17 we show the number of communities found by each algorithm on a 1000-vertex Erdös-Rényi random network with varying density. COPRA (with $v$ in the range 1,…,8) almost always finds a single community; the only exception is $v=8$ when the average degree is very low (plotted here). In contrast, the LFM algorithm finds about 200 communities, while CFinder finds a large and varying number of communities.

We have also compared the same four algorithms on our real networks. Table 2 shows the modularity and execution time of these algorithms. For CFinder, the modularity shown is the best of all values of its $k$ parameter. COPRA gives the best average modularity *and* execution time for every network tested, with the exception of "netscience", for which LFM does slightly better, and "word_association", for which it (like LFM) can find only a single "giant" community.

**Table 2.** Comparison of COPRA with other algorithms on real networks.

| Name | COPRA $Q_{ov}$ | CFinder $Q_{ov}$ | CONGO h=2 $Q_{ov}$ | CONGO h=3 $Q_{ov}$ | LFM $Q_{ov}$ | COPRA time (s) | CFinder time (s) | CONGO h=2 time (s) | CONGO h=3 time (s) | LFM time (s) |
|---|---|---|---|---|---|---|---|---|---|---|
| netscience | 0.812 | 0.606 | 0.760 | 0.802 | 0.861 | 0.04 | 0.25 | 1.3 | 1.5 | 4.0 |
| jazz | 0.716 | 0.596 | 0.338 | 0.519 | 0.641 | 0.03 | 16.9 | 25.4 | 53.2 | 4.5 |
| email | 0.506 | 0.463 | 0.244 | 0.260 | 0.098 | 0.22 | 4.00 | 30.6 | 617 | 39.6 |
| protein-protein | 0.552 | 0.456 | 0.377 | 0.381 | 0.169 | 0.75 | 2.75 | 8.6 | 96.5 | 101 |
| blogs | 0.748 | 0.376 | 0.497 | 0.524 | 0.508 | 2.9 | 3.05 | 6.5 | 33.8 | 121 |
| PGP | 0.788 | 0.568 | 0.563 | 0.579 | 0.318 | 14.4 | 34745 | 85.8 | 670 | 1499 |
| word_association | ~0 | 0.447 | 0.246 | 0.225 | 0.002 | – | 96.5 | 178 | 12026 | 1031 |
| blogs2 | 0.604 | 0.345 | 0.350 | 0.355 | 0.243 | 31.7 | 415 | 296 | 11702 | 17814 |
| cond-mat-2003 | 0.675 | 0.577 | 0.414 | 0.450 | 0.200 | 66.5 | 1134 | 1130 | 45110 | 13219 |

Finally, we experimented with a larger, weighted, network. "MathSciNet" [34] is a collaboration network whose giant component has 332689 vertices (authors) and 820644 weighted edges (coauthorship). The execution time is 8 minutes for COPRA and more than 20 hours for CFinder.

The interesting feature of this network is that a form of ground truth is available: each paper is tagged with one or more subject classes, so that each author can be categorized by the union of the classes associated with all of the author's papers. We *could* treat the subject classes as highly overlapping communities of authors: there are 6499 subject classes, of which each author belongs to 14 on average. However, it is unreasonable to expect these to coincide with network communities, because the authors in each class do not all collaborate with each other. Nevertheless, it is reasonable to assume that each network community should share a common subject class. If this is true, the success of a community detection algorithm can be measured by the fraction of authors in each community that have a subject class in common. We measured this fraction, which we call the *coverage*, for COPRA with $v=4$ and CFinder with $k=3$.

Figure 18 plots the number of communities of each size in logarithmic bins. With CFinder, some (low-degree) vertices are not included in any community, and so we plot these as singleton communities. For both COPRA and CFinder, the size of communities seems to vary according to a power law, but CFinder also produces one very large community. The figure also shows the coverage. Because this is influenced by the community size and the number of subject classes that each author has, we also plotted the expected coverage in the null model where there is no correlation between subjects and communities. We measured this by randomly swapping subject classes between authors millions of times while preserving the number of classes per author and the number of authors per class. Results show that (excluding the singleton communities and CFinder's giant community) the coverage is substantially better than expected by chance.

## 5. Bipartite networks

Many real networks are in a bipartite form: each vertex has one of two types (*modes*) such that edges connect only vertices of opposite modes. One example is a bibliographic network, in which vertices are papers and their authors. Community detection in a bipartite network can be done by *projecting* it onto each mode and finding communities in each of the projections, but this is usually less effective than detecting communities in the original bipartite network because information is lost in the projection [35]. Another disadvantage is that, in general, the communities detected in each projection are not mutually consistent. Better results can be obtained by detecting communities in the original bipartite network; several algorithms exist for this purpose, including those of Refs. [35–37].

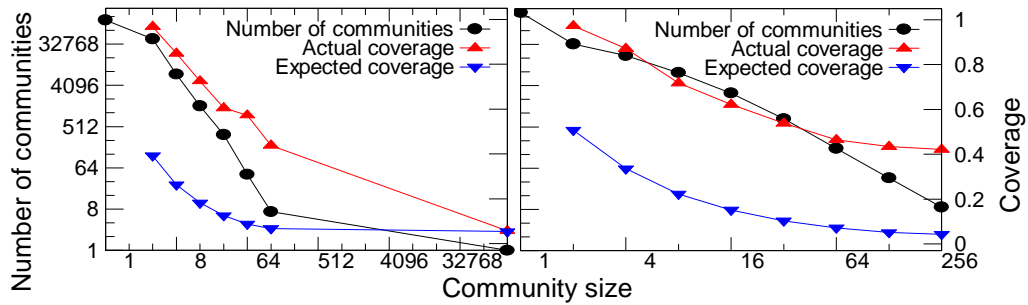

**Figure 18.** Size and coverage of communities detected in the "MathSciNet" network. The *coverage* of a community is the fraction of authors in it who share a common subject class. Results are plotted for bins such that bin 1 contains all communities with size 1 and bin $i$ ($i>1$) contains all communities with sizes in the range $2^{i-2}+1, \ldots, 2^{i-1}$. Left: CFinder, $k=3$. Right: COPRA, $v=4$.

Community detection algorithms for unipartite networks generally cannot be used for bipartite networks, and COPRA is no exception: community identifiers would never propagate from one set of vertices to the other. However, a simple change to the algorithm avoids this problem. Let $V_1$ and $V_2$ be the two sets of vertices, such that $V = V_1 \cup V_2$, $V_1 \cap V_2 = \emptyset$, and $|V_1| \geq |V_2|$. We change step 2 of our algorithm so that, in each iteration, vertex labels are propagated first from the larger (mode-1) set to mode 2, and then the *new* values of the mode-2 labels are propagated back to the mode-1 set:

> 2. For each vertex $x \in V_2$:
>     *Propagate*($x$, *old*, *new*).
>    For each vertex $x \in V_1$:
>     *Propagate*($x$, *new*, *new*).

Incidentally, the propagation of communities from one mode to the other is also a feature of the bipartite community detection algorithm of Barber [36].

We use two examples to show some results of bipartite community detection with COPRA. The first is the classic "Southern Women" network [38], which represents the attendance of 18 women at 14 events. We ran COPRA on this network 1000 times and measured the modularity of the "women" and "events" projections of each solution. (We do not use a bipartite modularity measure [35, 36] because these do not allow overlapping.) Results are shown in table 3. Although $v=2$ gives the best average modularity, the *maximum* modularity is similar for all $v$ up to 4.

**Table 3.** Modularity of COPRA on Southern Women bipartite network.

| $v$ | Mode 1: women | | | Mode 2: events | | |
| --- | --- | --- | --- | --- | --- | --- |
| | Max $Q_{ov}$ | Mean $Q_{ov}$ | SD $Q_{ov}$ | Max $Q_{ov}$ | Mean $Q_{ov}$ | SD $Q_{ov}$ |
| 1 | 0.404 | 0.278 | 0.164 | 0.530 | 0.386 | 0.222 |
| 2 | 0.401 | 0.313 | 0.146 | 0.530 | 0.429 | 0.198 |
| 3 | 0.404 | 0.286 | 0.176 | 0.534 | 0.384 | 0.237 |
| 4 | 0.404 | 0.174 | 0.194 | 0.534 | 0.228 | 0.255 |

Figure 19 shows the COPRA solution that corresponds to the maximum modularity, for each $v$: each row shows the two communities of women found. The communities in this network are not known, but COPRA's solution can be compared with the results of other methods, 21 of which are reviewed and analysed by Freeman [39]. All of our solutions contain exactly two communities, as do 20 of those in Ref. [39].

| $v=1$ | 1 2 3 4 5 6 7 9 | 8 10 11 12 13 14 15 16 17 18 |
|---|---|---|
| $v=2$ | 1 2 3 4 5 6 7 **9** | 8 **9** 10 11 12 13 14 15 16 17 18 |
| $v=3$ | 1 2 3 4 5 6 7 **8 9 10 16** | **8 9 10** 11 12 13 14 15 **16** 17 18 |
| $v=4$ | 1 2 3 4 5 6 7 **8 9 10 14 16** | **8 9 10** 11 12 13 **14** 15 **16** 17 18 |

**Figure 19.** Communities (of women) detected by COPRA in the Southern Women bipartite network.

Our solution for $v=1$ is identical to two of the solutions in Ref. [39], though most (17) solutions place woman 8 in the same community as women 1 to 7. Our solutions for $v \geq 2$ contain some overlapping, but only four of the solutions in Ref. [39] do. COPRA assigns woman 9 to both communities, as did the original study [38]. Woman 16 is placed in both communities by another of the solutions in Ref. [39], as well as COPRA.

Our second example is a bipartite bibliographic network of medical publications about autism. This is a connected network containing 4084 papers, 8498 authors, and 20099 edges. The unipartite projections onto papers and authors have 59935 and 51433 edges, respectively.

The best value of $v$, $v=9$, gives a solution with 372 communities that has modularities 0.615 and 0.720 for papers and authors, respectively. However, running COPRA on the projected network of papers finds (for $v=9$) a solution with 82 communities and a modularity of 0.816. In the network of authors (for $v=5$), a solution is found with 371 communities and a modularity of 0.793. They are shown in figure 20, together with the modularity found by CONGO for each of the projected networks for varying numbers of communities. The points for the bipartite results are circled.

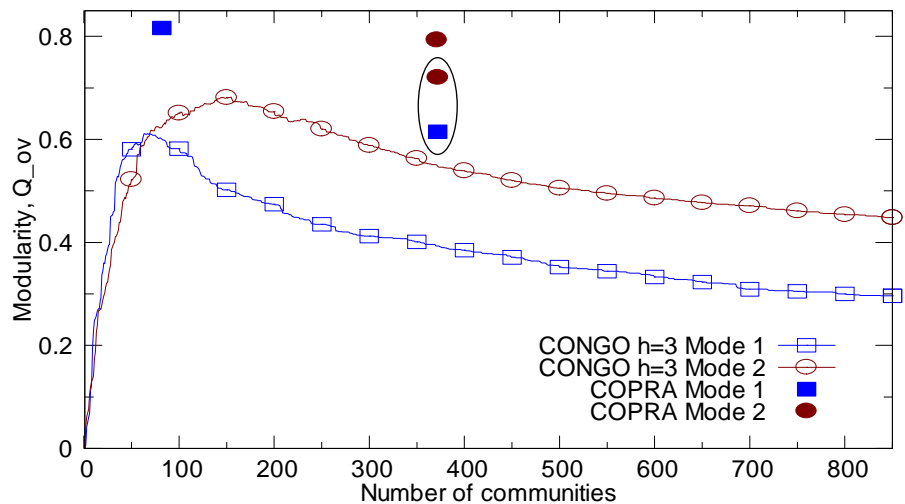

**Figure 20.** Modularity of bipartite (circled) and unipartite COPRA, and CONGO, on each mode of Autism bipartite network.

The COPRA results are all better (higher modularity) than CONGO's, but the bipartite communities have lower modularity than those found in the projections. The main reason for the low-modularity mode-1 solution is that modularity depends on the number of communities; the CONGO modularity also decreases between 82 and 372 communities. However, for mode 2, the number of communities in the bipartite and unipartite solutions are almost the same, but the modularity of the bipartite solution is lower. This is probably because there is a compromise between maximizing the modularity of each mode and maintaining the consistency between modes.

Our modified algorithm can also cope with "almost bipartite" networks, such as sexual networks. Provided we know the mode of every vertex, the algorithm works even if intramode edges are present. The RAK algorithm can also handle bipartite and "almost bipartite" networks without the need to distinguish between modes, because asynchronous updating allows community identifiers to propagate between modes.

## 6. Conclusions

### *6.1. Contributions*

We have presented an algorithm, COPRA, to detect overlapping communities in networks by label propagation. It is based on the RAK algorithm [14] but with non-trivial extensions that allow communities to overlap: we have extended the vertex label and the propagation step and introduced a novel termination condition that permits "synchronous updating". COPRA is guaranteed to terminate, and usually terminates with a good solution, as we have verified experimentally. We have also shown how our algorithm can handle weighted and bipartite networks.

The original RAK algorithm has been criticised, chiefly because of its tendency to find one of many suboptimal partitions [40, 41]. COPRA inherits some of these theoretical drawbacks, which we have not attempted to address in this paper. We merely note that many of the recent improvements to the RAK algorithm (e.g., [41, 42]) may also be applicable to COPRA.

### *6.2. Discussion*

Like the RAK algorithm, COPRA can be highly nondeterministic, with the community of a vertex often decided by a random choice. Nevertheless, on real networks, COPRA usually produces results that are better (in terms of modularity) than the other algorithms tested. This is not surprising in the light of recent work [43] showing that networks often have a vast number of high-modularity solutions, with no clear global maximum, and that these can differ fundamentally from each other. That is, even if COPRA (or any other algorithm) finds different solutions each time it is run, these solutions might all be good ones.

In our experiments on synthetic networks (figures 11-15), results are excellent and stable when there is relatively little mixing or overlap. However, when the mixing or overlap are increased too far, an incorrect random choice can cause a community identifier to propagate too far and flood the network, and performance suddenly declines. The LFM algorithm performs similarly to COPRA in this respect, while other algorithms (CFinder and CONGO) tend to give results that are worse but deteriorate more gradually as mixing or overlap increases.

COPRA performs better for larger networks and slightly better for smaller communities. In contrast, CFinder is strongly biased towards smaller communities but is unaffected by network size.

COPRA seems highly effective in discovering overlapping communities. The greater the value of $v$, the greater the overlap that can be detected by the algorithm. Moreover, a high value of $v$ does not impair the algorithm's ability to detect a smaller amount of overlap or even disjoint communities; for example, compare figures 11(c) and 11(d). However, if $v$ is increased too much, community identifiers start to propagate too far and performance drops (e.g., figure 6), especially for smaller networks. This is why we have only plotted $v=1,\ldots,6$ in figures 12-13 and $v=1,\ldots,8$ in figures 14-15: higher values of $v$ give worse results.

A potential user of COPRA might ask how the best value of $v$ should be chosen when trying to find communities in a real network. Essentially the same issue arises with any community detection algorithm that has a parameter (e.g., the number or density of communities). Our suggestion is to try the largest value of $v$ that gives a good (high modularity) solution. Running COPRA repeatedly, for different values of $v$, is often feasible because of its high speed.

Compared with the RAK algorithm, COPRA is less nondeterministic because it often retains several community identifiers for a vertex instead of making a random choice between them. Moreover, allowing $v$ to be greater than one reduces the number of iterations required, even though it increases the execution time per iteration (see figures 5 and 9).

Compared with other algorithms for overlapping community detection, the greatest advantage of COPRA is its speed. Overall execution time increases slightly more than linearly with the number of vertices but less than linearly with the average degree. In contrast, CFinder, LFM, and CONGO are all impractically slow for large or dense networks. This makes COPRA well-suited to finding overlapping communities in large real-world networks.

*6.3. Future work*

There are at least two directions in which this work could be extended. One is to make use of the belonging coefficients in the vertex labels to produce a "fuzzy" community assignment: each vertex would belong to community $c$ with strength $b$ if $(c,b)$ is in its vertex label. Another is to develop faster implementations. The algorithm is highly amenable to parallel implementation because each vertex can be updated independently during each propagation step, as a result of its use of synchronous updating.

Further information related to this paper, including the COPRA implementation and the network datasets used, can be found at www.cs.bris.ac.uk/~steve/networks/ .

**Acknowledgments**

Thanks are due to Lnx Research for supplying the Autism bibliographic dataset used in section 5, and to the referees for helpful comments.

**References**

[1] Girvan M and Newman M E J 2002 Community structure in social and biological networks *Proc. Natl. Acad. Sci. USA* **99** 7821
[2] Newman M E J 2004 Fast algorithm for detecting community structure in networks *Phys. Rev.* E **69** 066133
[3] Palla G, Derényi I, Farkas I and Vicsek T 2005 Uncovering the overlapping community structure of complex networks in nature and society *Nature* **435** 814
[4] Hofman J M and Wiggins C H 2008 Bayesian approach to network modularity *Phys. Rev. Lett.* **100** 258701
[5] Fortunato S 2010 Community detection in graphs *Phys. Rep.* **486** 75
[6] Alba R D 1973 A graph-theoretic definition of a sociometric clique *J. Math. Sociol.* **3** 113
[7] Baumes J, Goldberg M and Magdon-Ismail M 2005 Efficient identification of overlapping communities *Lect. Notes Comput. Sc.* **3495** 27
[8] Gregory S 2007 An algorithm to find overlapping community structure in networks *Lect. Notes Comput. Sci.* **4702** 91
[9] Gregory S 2008 A fast algorithm to find overlapping communities in networks *Lect. Notes Comput. Sci.* **5211** 408
[10] Zhang S, Wang R and Zhang X 2007 Identification of overlapping community structure in complex networks using fuzzy C-means clustering *Physica* A **374** 483
[11] Blondel V D, Guillaume J-L, Lambiotte R and Lefebvre E 2008 Fast unfolding of communities in large networks *J. Stat. Mech.* P10008
[12] Clauset A, Moore C and Newman M E J 2008 Hierarchical structure and the prediction of missing links in networks *Nature* **453** 98
[13] Lancichinetti A, Fortunato S and Kertész J 2009 Detecting the overlapping and hierarchical community structure of complex networks *New J. Phys.* **11** 033015
[14] Raghavan U N, Albert R and Kumara S 2007 Near linear time algorithm to detect community structures in large-scale networks *Phys. Rev.* E **76** 036106
[15] Leung I X Y, Hui P, Liò P and Crowcroft J 2009 Towards real-time community detection in large networks *Phys. Rev.* E **79** 066107
[16] Rand W M 1971 Objective criteria for the evaluation of clustering methods *J. Am. Stat. Assoc.* **66** 846
[17] Hubert L and Arabie P 1985 Comparing partitions *J. Classif.* **2** 193
[18] Fred A L N and Jain A K 2003 Robust data clustering *Proc. IEEE Computer Society Conf. on Computer Vision and Pattern Recognition* pp 128-133
[19] Collins L M and Dent C W 1988 Omega: A general formulation of the Rand index of cluster recovery suitable for non-disjoint solutions *Multivar. Behav. Res.* **23** 231
[20] Newman M E J and Girvan M 2004 Finding and evaluating community structure in networks *Phys. Rev.* E **69** 026113
[21] Newman M E J 2006 Modularity and community structure in networks *Proc. Natl. Acad. Sci. USA* **103** 8577

[22] Nicosia V, Mangioni G, Carchiolo V and Malgeri M 2009 Extending the definition of modularity to directed graphs with overlapping communities *J. Stat. Mech.* P03024
[23] Nicosia V, Mangioni G, Carchiolo V and Malgeri M 2008 Extending modularity definition for directed graphs with overlapping communities arXiv:0801.1647
[24] Adamcsek B, Palla G, Farkas I, Derényi I and Vicsek T 2006 CFinder: Locating cliques and overlapping modules in biological networks *Bioinformatics* **22** 1021
[25] Lancichinetti A, Fortunato S and Radicchi F 2008 Benchmark graphs for testing community detection algorithms *Phys. Rev.* E **78** 046110
[26] Lancichinetti A and Fortunato S 2009 Community detection algorithms: a comparative analysis *Phys. Rev.* E **80** 056117
[27] Newman M E J 2006 Finding community structure in networks using the eigenvectors of matrices *Phys. Rev.* E **74** 036104
[28] Gleiser P and Danon L 2003 Community structure in jazz *Adv. Complex Syst.* **6** 565
[29] Guimerà R, Danon L, Diaz-Guilera A, Giralt F and Arenas A 2003 Self-similar community structure in a network of human interactions *Phys. Rev.* E **68** 065103(R)
[30] Boguña M, Pastor-Satorras R, Diaz-Guilera A and Arenas A 2004 Models of social networks based on social distance attachment *Phys. Rev.* E **70** 056122
[31] Newman M E J 2001 The structure of scientific collaboration networks *Proc. Natl. Acad. Sci. USA* **98** 404
[32] http://www.cfinder.org
[33] www.cs.bris.ac.uk/~steve/networks/software/conga.html
[34] Palla G, Farkas I, Pollner P, Derényi I and Vicsek T 2008 Fundamental statistical features and self-similar properties of tagged networks *New J. Phys.* **10** 123026
[35] Guimerà R, Sales-Pardo M and Amaral L A N 2007 Module identification in bipartite and directed networks *Phys. Rev.* E **76** 036102
[36] Barber M J 2007 Modularity and community detection in bipartite networks *Phys. Rev.* E **76** 066102
[37] Du N, Wang B, Wu B and Wang Y 2008 Overlapping community detection in bipartite networks *Proc. IEEE/WIC/ACM International Conference on Web Intelligence and Intelligent Agent Technology* p 176
[38] Davis A, Gardner B B and Gardner M R 1941 *Deep South* (Chicago: University of Chicago Press)
[39] Freeman L C 2003 Finding social groups: a meta-analysis of the Southern Women data *Dynamic Social Network Modeling and Analysis* ed R Breiger, K Carley and P Pattison (Washington, DC: The National Academies Press) pp 39-77
[40] Tibély G and Kertész J 2008 On the equivalence of the label propagation method of community detection and a Potts model approach *Physica* A **387** 4982
[41] Barber M J and Clark J W 2009 Detecting network communities by propagating labels under constraints *Phys. Rev.* E **80** 026129
[42] Liu X and Murata T 2010 Advanced modularity-specialized label propagation algorithm for detecting communities in networks *Physica* A **389** 1493
[43] Good B H, de Montjoye Y-A and Clauset A 2010 The performance of modularity maximization in practical contexts *Phys. Rev.* E **81** 046106